\documentclass[reprint,secnumarabic,amssymb, nobibnotes, aps, prl,superscriptaddress,showpacs]{revtex4-1}

\usepackage{graphics}
\usepackage{graphicx}
\usepackage{hyperref}
\usepackage{amsmath}
\usepackage{bm}

\def\gb{$\bar{\Gamma}$\ }

\def\mb{$\bar{\mathrm{M}}$\ }

\def\gmb{${\bar{\Gamma}}{\bar{\mathrm{M}}}$\ }

\def\gkb{${\bar{\Gamma}}{\bar{\mathrm{K}}}$\ }

\def\EF{$E_{\mathrm{F}}$\ }
\def\EFc{$E_{\mathrm{F}}$, }

\begin{document}

\title{Alkali-metal induced band structure deformation investigated by angle-resolved photoemission spectroscopy and first-principles calculations}
\author{S. Ito}
\affiliation{Institute for Solid State Physics (ISSP), The University of Tokyo, Kashiwa, Chiba 277-8581, Japan}
\author{B. Feng}
\affiliation{Hiroshima Synchrotron Radiation Center (HSRC), Hiroshima University, 2-313 Kagamiyama, Higashi-Hiroshima 739-0046, Japan}
\author{M. Arita}
\affiliation{Hiroshima Synchrotron Radiation Center (HSRC), Hiroshima University, 2-313 Kagamiyama, Higashi-Hiroshima 739-0046, Japan}
\author{T. Someya}
\affiliation{Institute for Solid State Physics (ISSP), The University of Tokyo, Kashiwa, Chiba 277-8581, Japan}
\author{W.-C. Chen}
\affiliation{National Synchrotron Radiation Center (NSRRC), Hsinchu, Taiwan 30076, Republic of China}
\author{A. Takayama}
\affiliation{Department of Physics, The University of Tokyo, Bunkyo-ku, Tokyo 113-0033, Japan}
\author{T. Iimori}
\affiliation{Institute for Solid State Physics (ISSP), The University of Tokyo, Kashiwa, Chiba 277-8581, Japan}
\author{H. Namatame}
\affiliation{Hiroshima Synchrotron Radiation Center (HSRC), Hiroshima University, 2-313 Kagamiyama, Higashi-Hiroshima 739-0046, Japan}
\author{M. Taniguchi}
\affiliation{Hiroshima Synchrotron Radiation Center (HSRC), Hiroshima University, 2-313 Kagamiyama, Higashi-Hiroshima 739-0046, Japan}
\author{C.-M. Cheng}
\affiliation{National Synchrotron Radiation Center (NSRRC), Hsinchu, Taiwan 30076, Republic of China}
\author{S.-J. Tang}
\affiliation{National Synchrotron Radiation Center (NSRRC), Hsinchu, Taiwan 30076, Republic of China}
\affiliation{Department of Physics and Astronomy, National Tsing Hua University, Hsinchu, Taiwan 30013, Republic of China}
\author{F. Komori}
\affiliation{Institute for Solid State Physics (ISSP), The University of Tokyo, Kashiwa, Chiba 277-8581, Japan}
\author{I. Matsuda}
\affiliation{Institute for Solid State Physics (ISSP), The University of Tokyo, Kashiwa, Chiba 277-8581, Japan}
\date{\today}
\begin{abstract}

Alkali-metal adsorption on the surface of materials is widely used for \textit{in situ} surface electron doping, particularly for observing unoccupied band structures by angle-resolved photoemission spectroscopy (ARPES). However, the effects of alkali-metal atoms on the resulting band structures have yet to be fully investigated, owing to difficulties in both experiments and calculations. Here, we combine ARPES measurements on cesium-adsorbed ultrathin bismuth films with first-principles calculations of the electronic charge densities and demonstrate a simple method to evaluate alkali-metal induced band deformation. We reveal that deformation of bismuth surface bands is directly correlated with vertical charge-density profiles at each electronic state of bismuth. In contrast, a change in the quantized bulk bands is well described by a conventional rigid-band-shift picture. We discuss these two aspects of the band deformation holistically, considering spatial distributions of the electronic states and cesium-bismuth hybridization, and provide a prescription for applying alkali-metal adsorption to a wide range of materials.
\end{abstract}
\maketitle 

\section{I. Introduction}
Owing to the large electronegativity difference, alkali-metal atoms adsorbed on the surface of materials are easily ionized and give up their valence electrons to the substrate atoms \cite{kiejna1981,aruga1989,stampfl1994, diehl1997}. Recently, in combination with rapid development of angle-resolved photoemission spectroscopy (ARPES), this \textit{in situ} surface electron doping has been widely used to observe unoccupied band structures, which cannot be accessed by ARPES alone \cite{zhu2011,zhang2014,zhu2015,feng2016}. This method is also an important tool to flexibly tune carrier concentration of substrates \cite{ohta2006, kim2015,miyata2015,ren2017}. In most of these applications, it is implicitly assumed that alkali-metal adsorption does not seriously deform the original band structures, that is, a rigid-band-shift picture is considered.

However, this picture cannot always be valid because introducing alkali-metal atoms can inherently modify the original system. Theoretical calculations showed that adsorbed alkali-metal atoms localize at particular adsorption sites on the surface plane \cite{stampfl1994, diehl1997}, which can selectively affect specific orbitals of substrate atoms. Very recently, such effect was suggested to occur by first-principles calculations of potassium-adsorbed black phosphorus, where the valence band top and the conduction band bottom showed different amounts of shifts \cite{kim2017}. Recent demand for precise determination of unoccupied electronic structures in novel materials, on the energy scale of dozens of meV \cite{zhu2015,feng2016, belopolski2016}, calls for a detailed understanding and evaluation of the alkali-metal induced band deformation. However, the effect on fine electronic structures across wide energy and momentum ranges has not yet been comprehensively studied. A major difficulty in experiments is the requirement for precise sample alignment to detect small modifications in the band structures. More seriously, to directly address this problem with first-principles calculations, a very large supercell is required to express the low coverage ($0.01\sim0.1$ML) used in experiments, which results in tremendous computational costs. To make alkali-metal adsorption a handy tool for determining unoccupied electronic structures, even at smaller energy scales, methods to evaluate the additional effects should be much simpler.

Very recently, a strong deformation of the surface-state bands due to cesium (Cs) adsorption was observed across a wide momentum range on an ultrathin bismuth (Bi) film  \cite{matetskiy2017}; however, the underlying mechanism is not yet fully understood. Because the surface bands of Bi are complexly hybridized with the bulk bands \cite{hofmann2006}, the observed behavior can be related to a spatial distribution of each state in the surface-normal direction. Thanks to the strong quantum-size effect in ultrathin Bi films, quantized bulk states, known as quantum-well-states (QWSs), can visualize a broad background of bulk bands as sharp peaks in ARPES \cite{chiang2000,hirahara2007,bian2009,takayama2012,ito2016}. QWS bands also have direct correspondence to those obtained in a slab calculation used for examining surface effects. From a technical point of view, the extremely anisotropic shape of Bi Fermi surfaces facilitates precise alignment of the sample orientations \cite{hofmann2006, ohtsubo2012}. All these characteristics make ultrathin Bi films an ideal platform for investigating alkali-metal induced effects on band structures.

In the present paper, we combined ARPES measurements on Cs-adsorbed ultrathin Bi films with first-principles calculations of the electronic charge distributions inside a pristine Bi slab as a simple approach to evaluate the alkali-metal induced band structure deformation. We experimentally extracted the modification of Bi band structures caused by Cs adsorption and directly compared them with calculations of the charge density distributions in bare Bi. This effective combination between conventional methods enables us to associate band deformation with local charge distributions, avoiding the difficulties discussed above. First we revealed that the deformation of Bi surface bands was strongly correlated with vertical charge density profiles across the Bi film. The QWS (bulk) bands exhibited, however, a rigid-shift behavior contrary to the prediction based on the charge profiles. We interpreted the different behaviors based on their spatial distributions against that of Cs-Bi hybridization. We also introduced another picture of the band deformation using an effective model analysis from the viewpoint of modulation in a surface confinement potential.

\section{II. Methods}
\begin{figure*}
\includegraphics{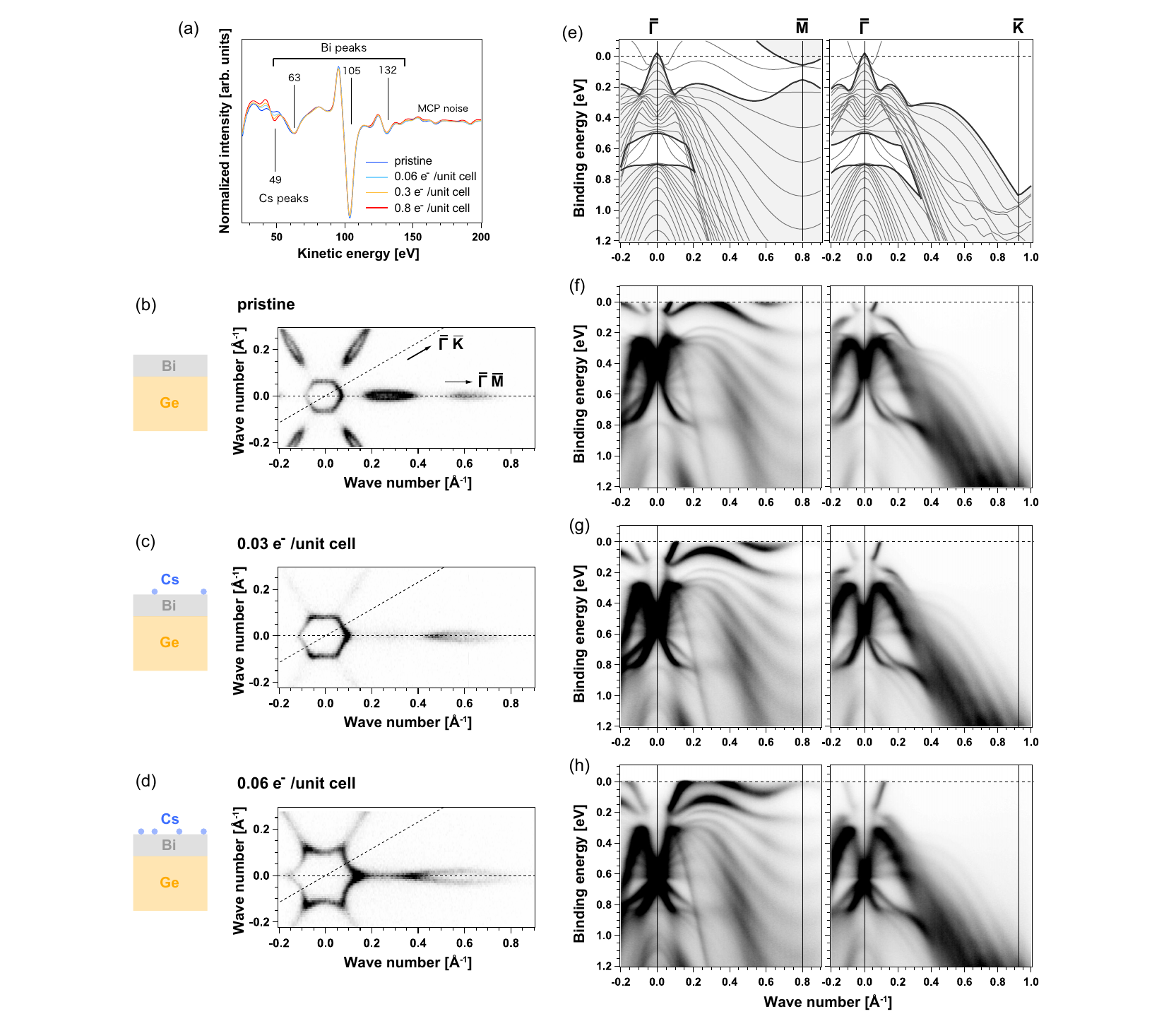}
\caption{\label{fig1}(a) Auger electron spectra with increasing Cs coverage. Cs coverage is expressed as the number of electrons doped per unit cell, as calibrated from the change of Fermi surface areas. (b)-(d) Fermi surfaces extracted with an energy window of 0.01 eV on a pristine Bi(111) film and Cs-adsorbed films with two different coverages. Schematics of the samples and two high-symmetry directions are illustrated. (e) Band structures calculated along \gmb and \gkb directions on a pristine Bi slab. Shaded areas depict bulk projections. (f)-(h) Band structures measured along \gmb and \gkb directions on a Bi film with Cs coverages corresponding to those in (b)-(d).}
\end{figure*}

An ultrathin Bi(111) film was grown on a Ge wafer cut in the [111] direction and cleaned by cycles of Ar$^+$ sputtering and annealing at 900 K. Evaporation of Bi was performed at room temperature and followed by annealing at 400 K \cite{hatta2009}. The quality of the substrate and the film was confirmed from low-energy electron diffraction measurements. The film thickness was calibrated as 14 BL by comparing the QWS energy splitting with previous reports \cite{hirahara2007,bian2009,takayama2012,ito2016}. Cs atoms were adsorbed onto the film at $\sim$200 K by commercial dispensers (SAES Getters S.p.A.). The purity of the Cs source was checked by Auger electron spectroscopy \cite{davis1976}. Figure \ref{fig1}(a) shows the Auger spectra for different Cs coverages, where the coverage is expressed as the number of electrons doped into a Bi film as described later. We can confirm only the Cs peak intensity gradually increased. ARPES measurements were performed at BL-9A of HSRC and BL-21B1 of NSRRC. The measurement photon energy and temperature were set to 21 eV and 20 K, respectively, and the total energy resolution was 12 meV. In both experimental stations, an automated six-axis orientation controller was used, which enabled precise sample alignment, as shown in Figs. \ref{fig1}(b)-(d).

First-principles calculations were performed using the ABINIT code \cite{gonze2002}. The Perdew-Burke-Ernzerhof generalized gradient approximation (GGA-PBE) was used to describe the exchange-correlation functional \cite{perdew1996}. A Hartwigsen-Goedecker-Hutter norm-conserving pseudopotential was used, in which spin-orbit coupling was implemented \cite{hartwigsen1998}. A free-standing Bi slab was used and the length of a vacuum region was set to 5 BL ($\sim$19.5 \AA). Lattice constants of the slab were fixed to the experimental values \cite{cucka1962,liu1995}. All the calculations in this paper were performed on both 14 BL and 15 BL slabs and the figures in this paper are based on the results from the latter, which featured QWS energy levels closer to experimental ones. We confirmed that the conclusions did not depend on a slab thickness. A Monkhorst-Pack grid for $k$-point sampling as 7$\times$7$\times$1 was selected \cite{monkhorst1976}. For the calculation of the bulk projections, a rhombohedral unit cell and 9$\times$9$\times$9 $k$-point sampling were used. Convergence against all the important parameters was confirmed. 

\section{III. Results and discussion}

\subsection{A. Overview of Cs-induced effects}
Figures \ref{fig1}(b)-\ref{fig1}(d) shows the Fermi surfaces measured on a (111) surface of a pristine Bi film and a Cs-adsorbed film with two different Cs coverages. Owing to electrons doped from the adsorbates, the areas corresponding to the electron and the hole pockets gradually increased and decreased, respectively. From this modulation of the Fermi surfaces, we calibrated the number of electrons doped per unit cell \cite{crain2005}. Figures \ref{fig1}(e) and \ref{fig1}(f) respectively show the calculated and measured band structures of a pristine Bi film. Shaded areas in Fig. \ref{fig1}(e) depict regions of bulk projections. Surface bands outside the bulk projections were observed with strong intensities in Fig. \ref{fig1}(f). QWS bands inside the bulk projections were also sharply observed. In the ARPES images after Cs adsorption in Figs. \ref{fig1}(g) and \ref{fig1}(h), electron doping from the Cs atoms caused shifting of the overall band structures toward higher binding energy. However, as is consistent with the previous paper \cite{matetskiy2017}, the variation in the degree of band shifting at some ($E, k$) points resulted in deformation of the band structures, particularly around the surface bands located near \EF and 0.6$\sim$0.8 eV. In contrast, the QWS bands appear to show a rigid-shift behavior.

\begin{figure*}
\includegraphics{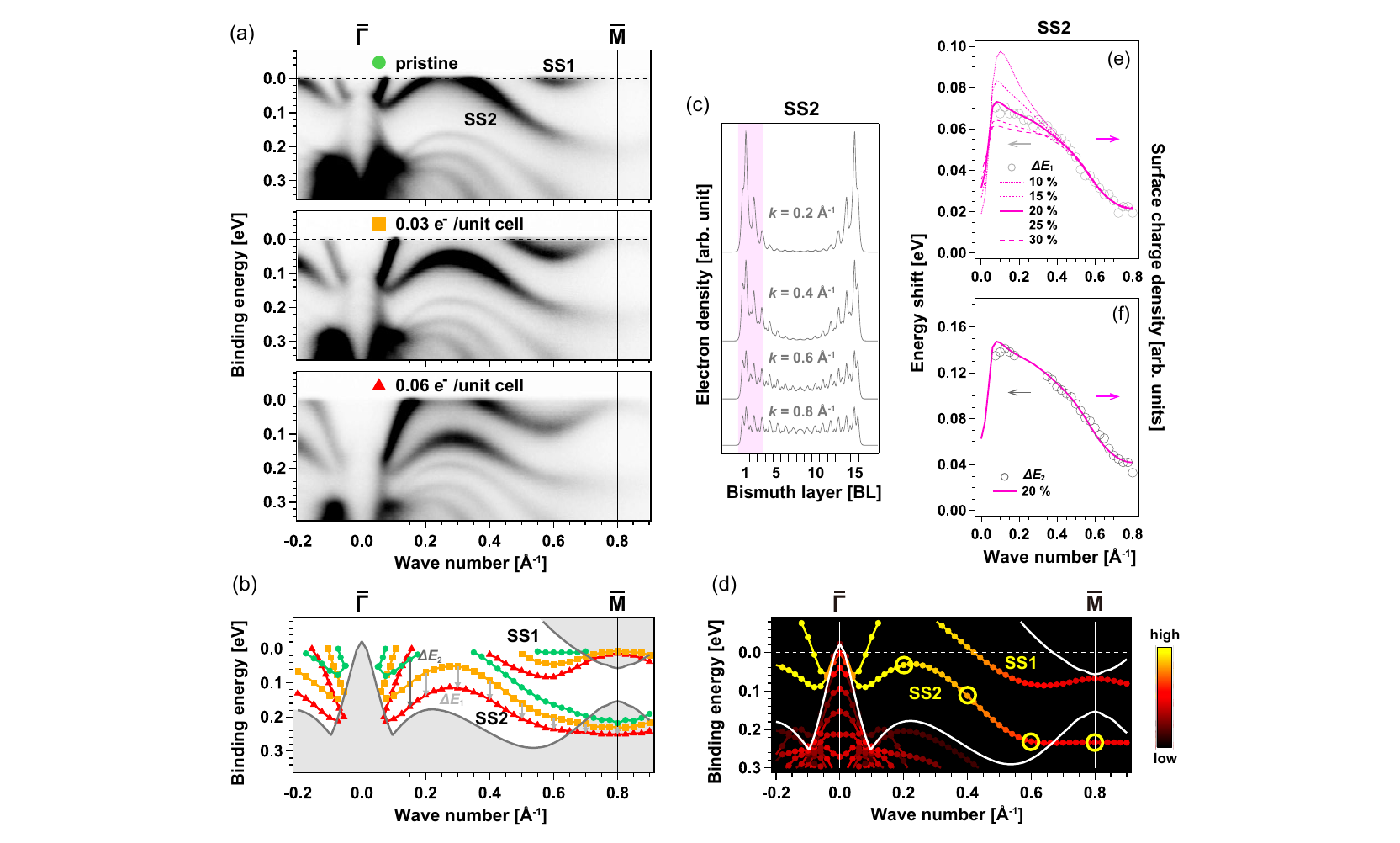}
\caption{\label{fig2}(a) ARPES images magnified near \EF along \gmb direction with increasing Cs coverage. (b) Peak positions of the SS1 and SS2 bands at each coverage extracted with Lorentzian fitting. (c) Calculated charge distributions along the out-of-plane direction at ($E, k$) points highlighted in (d). (d) Calculated band structures and surface charge densities mapped with a color scale. The grey and white solid lines in (b) and (d) depict the calculated bulk projections. (e) Comparison between wavenumber-dependence of surface charge densities calculated with several ratios and that of the SS2 band shifts between two coverages of 0.03e$^-$ and 0.06 e$^-$ per unit cell, $\Delta E_1$. Surface charge densities are superimposed with an arbitrary scale. (f) Same as (e) for the SS2 band shifts between a pristine and a 0.06 e$^-$ cases, $\Delta E_2$.}	
\end{figure*}

\subsection{B. Deformation of Bi surface bands}
First we analyzed the modifications of the surface bands in detail. Figure \ref{fig2}(a) shows the ARPES images magnified around the two surface bands near \EFc SS1 and SS2, along \gmb direction with increasing Cs coverages. We extracted the peak positions of the SS1 and SS2 bands with Lorentzian fitting and superimposed the peak positions with different coverages in Fig. \ref{fig2}(b), where the deformation of the surface bands becomes more obvious. (The raw spectra are presented in Appendix A.) The degree of shifting is smaller around the regions overlapping with bulk projections near \mb and becomes greater away from the bulk edges, which likely reflects the surface-like character of each electronic state in Bi. To test this hypothesis, we calculated the electronic charge distributions along the out-of-plane direction, where the in-plane distributions were integrated. Figure \ref{fig2}(c) shows the vertical charge density profiles calculated at four $k$ points on SS2 highlighted in Fig. \ref{fig2}(d). At $k=0.2$ \AA$^{-1}$, where the SS2 band is located inside a band gap, the corresponding charge distribution is localized near the surface layers. In contrast, the distribution at $k=0.8$ \AA$^{-1}$, where the band overlaps a bulk projection, has a finite intensity in the middle layers and exhibits bulklike character. 

As a next step we integrated the charge densities near the surface layers. This quantity reflects how localized around the surface each electronic state is, as the total charge density is normalized for every state in the present calculation. Because there is no general standard for a surfacelike state, we calculated the surface charge density using several ``surface length ratio'' against the total thickness. For example, the shaded area in Fig. \ref{fig2}(c) shows a region confined to a depth of 20\% of the total thickness. We mapped the obtained surface charge densities on band structures with a color scale in Fig. \ref{fig2}(d). The increasing tendency for the SS2 band shifts in Fig. 2(b) and that of the surface charge densities in Fig. 2(d) correspond well with each other. To quantitatively confirm this correspondence, we superimposed the wave-number dependence of the surface charge densities extracted with several surface length ratios on that of the SS2 band shifts for two coverages of 0.03 e$^-$ and 0.06 e$^-$ per unit cell, as shown in Fig. \ref{fig2}(e). In the case of the length ratio of 20 \%, these two showed an excellent agreement. Here, the surface densities were plotted with arbitrary units and the agreement was manually adjusted in terms of the absolute values. Nevertheless, the reproducibility of the shape of the experimental wave-number dependence is still remarkable. We also performed the same analysis on the band shifts for pristine and 0.06 e$^-$ cases, which was also nicely reproduced as shown in Fig. \ref{fig2}(f).

\begin{figure}
\includegraphics{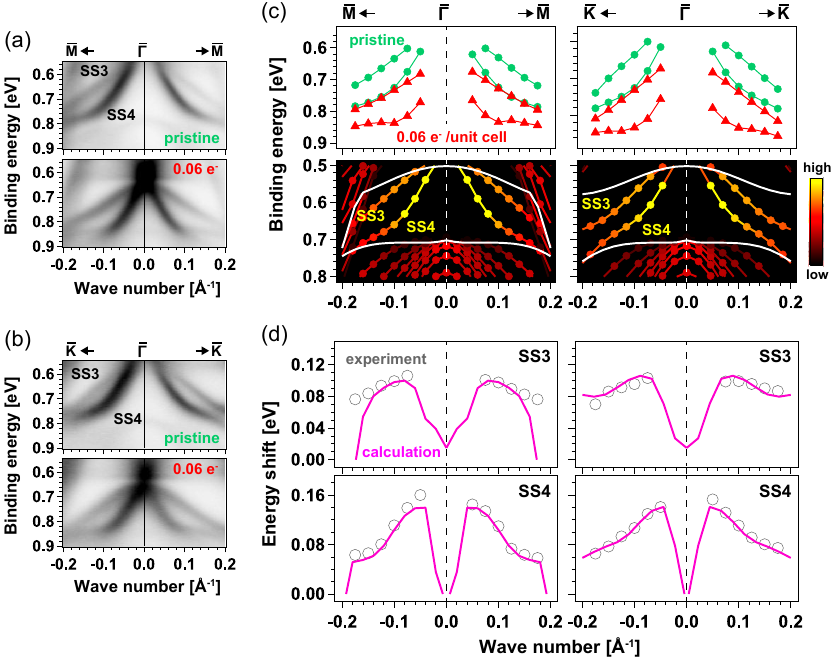}
\caption{\label{fig3}(a), (b) ARPES images magnified around the SS3 and SS4 bands between a pristine and a 0.06 e$^-$ cases in \gmb and \gkb directions, respectively. (c) Extracted peak positions and calculated band structures with surface charge densities mapped. (d) Comparison of experimental and calculated band shifts. The latter was converted from the surface charge densities using a linear scaling obtained in Fig. \ref{fig2}(f).}	
\end{figure}

We further tested the correspondence using the surface bands located at 0.6$\sim$0.8 eV around \gb point, SS3 and SS4. Figures \ref{fig3}(a) and \ref{fig3}(b) show ARPES images magnified around these bands for pristine and 0.06 e$^-$ cases. In the same manner as in Fig. \ref{fig2}, we extracted the peak positions and calculated the surface charge densities as shown in Fig. \ref{fig3}(c). We converted the surface densities to band shift values using a linear scaling relation obtained in Fig. \ref{fig2}(f), and compared the experimental and the calculated band shifts in Fig. \ref{fig3}(d). They showed an excellent agreement both qualitatively and quantitatively. The surprising applicability for electronic states located in wide energy and momentum ranges strongly supports the connection between band shifts and surface charge densities.

\subsection{C. Change in the quantized bulk bands}

\begin{figure}
\includegraphics{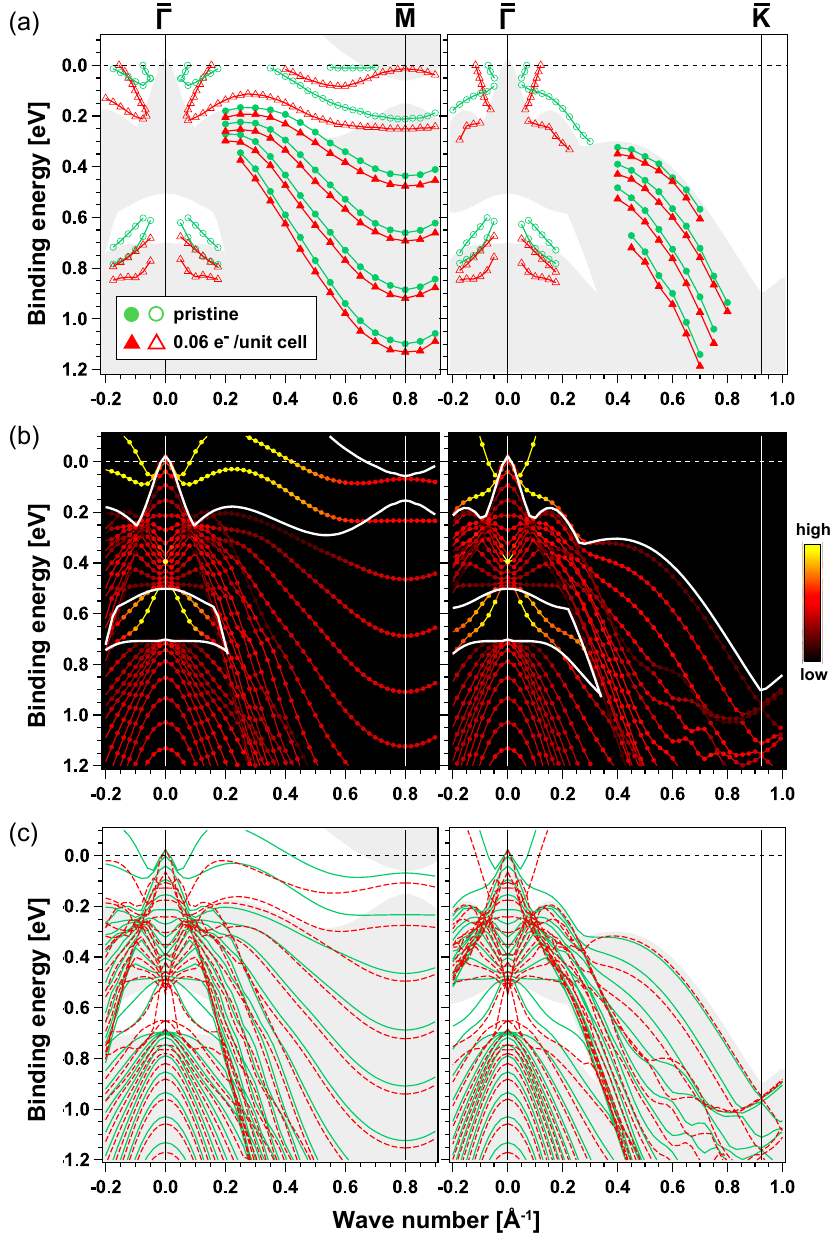}
\caption{\label{fig4} (a) Extracted peak positions of surface (unfilled marker) and QWS (filled maker) bands in pristine and 0.06 e$^-$ cases along \gmb and \gkb directions. (b) Band structures and surface charge densities calculated in the same manner as in Fig. \ref{fig2}(d) with a wider energy range. (c) Calculated band structures without (green solid line) and with (red dashed line) energy shifts converted from the surface charge densities as done in Fig. \ref{fig3}(d).} 	
\end{figure}

To further check the correspondence revealed in Section B, we applied the analysis to the QWS bands. Figure \ref{fig4}(a) shows the peak positions extracted by Lorentzian fitting for the QWS bands of pristine and 0.06 e$^-$ cases. The results of the surface bands are shown together. Figure \ref{fig4}(b) depicts the wide-range band structures and the surface charge densities calculated in the same manner as in Fig. \ref{fig2}(d). Furthermore, as we performed in Fig. \ref{fig3}(d), we converted the surface densities to energy shifts of each state and simulated the band structures before and after applying the shifts in Fig. \ref{fig4}(c). 

Whereas an almost perfectly rigid-shift behavior was observed for the QWS bands in Fig. \ref{fig4}(a), the band shifts predicted from the surface charge densities were clearly dependent on the wave number, as shown in Fig. \ref{fig4}(c). Nevertheless, the excellent reproducibility in the surface band shifts still seems to support the empirical correlation between the band deformation and the surface charge densities. We interpreted the observed deviation as a signature of effectively different mechanisms between surface and QWS band shifts, which are discussed in the next section.

\subsection{D. Mechanisms of the band deformation}
\begin{figure}
\includegraphics{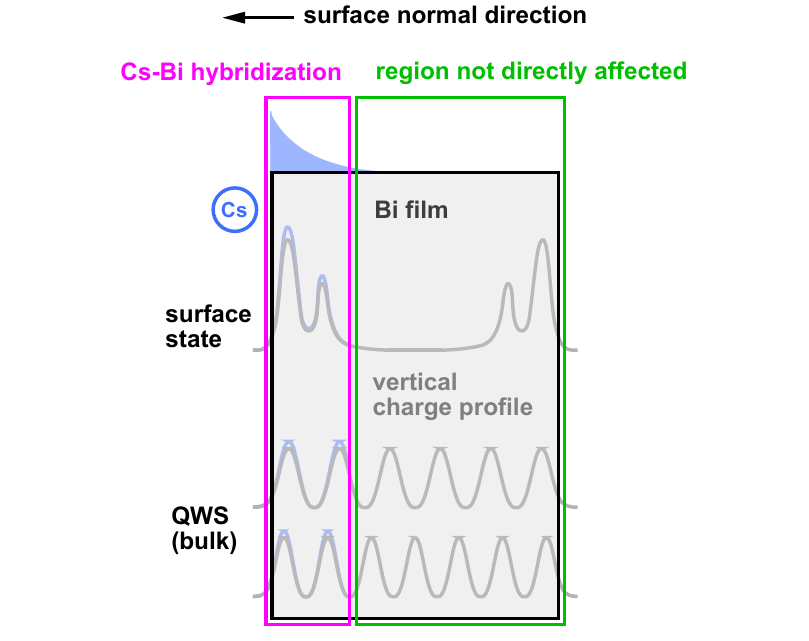}
\caption{\label{fig5}Schematic drawing of the mechanisms of the band deformation due to Cs adsorption on the surface of a Bi film. The effects of Cs are depicted using blue color.} 	
\end{figure}

The essence of the Cs adsorption is a formation of Cs-Bi bonding, whose effect rapidly decays away from the surface. Figure \ref{fig5} shows a schematic drawing of spatial distributions for the Cs-Bi hybridization and the charge densities of each electronic state in Bi. The surface state, whose vertical charge-density profile is localized around this region, is sensitive to the effect and the degree of the modulation is determined by the extent of the localization. In contrast, the QWS has a vertical charge-density profile spread almost uniformly throughout the film and a major part of the charge profile is not seriously affected by the Cs-Bi hybridization. As a result, the band modulation can be well described as a change in the filling of electrons, that is, the chemical potential.

Strictly speaking, it is impossible to exactly distinguish the two cases and the band deformation must be described as a renewal of the overall band structure before and after introducing Cs atoms into the system. Such perspective of alkali-metal adsorption is complete but can only be achieved by first-principles calculations directly handling adsorbates. What we demonstrated here is that the resulting band structure can be nicely approximated by incorporating the band deformation correlated with the surface charge densities into a conventional rigid-band-shift model.

Finally, we point out that effects of alkali-metal adsorption have also been studied from the viewpoint of a surface confinement potential \cite{kiejna1981,aruga1989,stampfl1994, diehl1997}. In Appendix B, we demonstrated that the perspective provides another interpretation of the surface band deformation based on changes in the substrate's work function and the phase shift depending on the band structure. This approach is also applicable to QWSs and may lead to deeper understanding of the mechanisms in combination with further experiments systematically controlling both Cs coverages and Bi thicknesses.

\section{IV. Conclusion}
We presented a simple approach to evaluate alkali-metal induced band structure deformation by combining ARPES measurements on Cs-adsorbed ultrathin Bi films with first-principles calculations of electron charge distributions inside a pristine Bi slab. Here Cs-induced modifications of Bi band structures were extracted from ARPES measurements and information on electronic charge distributions at corresponding electronic states was obtained from calculations on bare Bi. We revealed the deformation of Bi surface bands was directly connected to the vertical charge density profiles across the Bi film, whereas the quantized bulk bands showed a rigid-shift behavior regardless of the charge profiles. We attributed the different behaviors to their sensitivity to Cs-Bi hybridization near the surface depending on the spatial distribution of each state. The present paper provides a prescription to facilitate the usage of alkali-metal adsorption for exploring fine electronic structures in a wide range of materials.

\section{ACKNOWLEDGEMENTS}
We acknowledge G. Bian and K. Kobayashi for advice on first-principle calculations and Y. Endo for discussions on alkali-metal adsorption. We also thank A. Jackson and M. Cameau for checking the manuscript. The ARPES measurements were performed with the approval of the Proposal Assessing Committee of HSRC (Proposal No. 15-A-38) and the Proposal Assessing Committee of NSRRC (Project No. 2015-2-090-1). S.I. acknowledges support by JSPS under KAKENHI Grant No. 17J03534. S.I. was also supported by JSPS through the Program for Leading Graduate Schools (ALPS).

\section{Appendix A: Raw photoemission spectra}

\begin{figure*}
\includegraphics{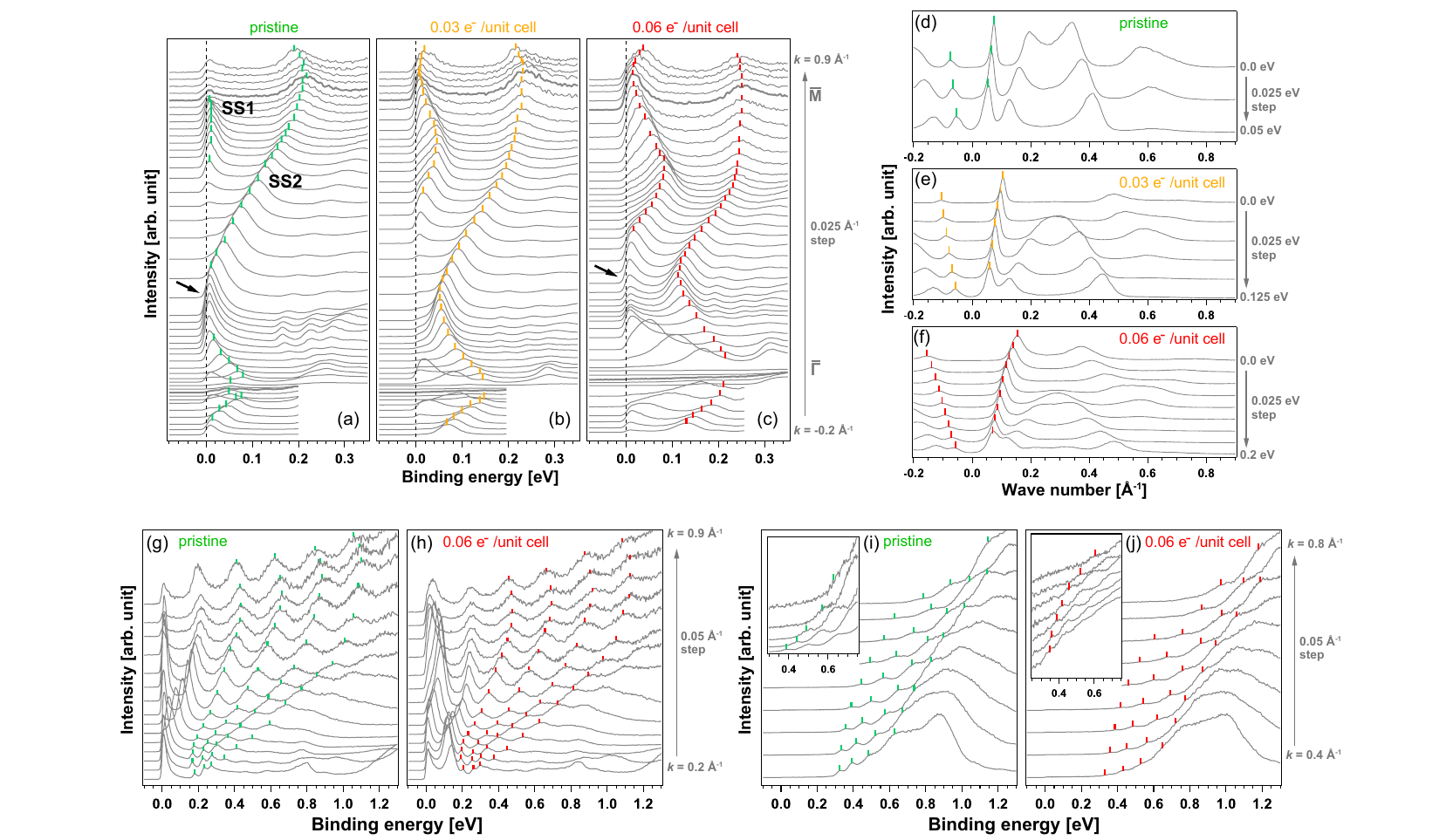}
\caption{\label{fig6}(a)-(c) Energy distribution curves and (d)-(f) momentum distribution curves extracted from the ARPES images used in Fig. \ref{fig2}. The curves were manually scaled and offset for the clear visualization. The wave number and energy values are summarized in the right axes. Peak positions determined by Lorentzian fitting are shown with markers. The dashed lines describe the Fermi level and the thick solid lines indicate EDCs at high-symmetry points (\gb and $\bar{\mathrm{M}}$). The black arrows highlight specific peaks discussed in the main text. (g), (h) Same as (a)-(c) but for a wider energy range to show QWS peak positions used in Fig. \ref{fig4}(a) along \gmb direction. (i), (j) Same as (g) and (h) but for QWSs along \gkb direction. The insets magnify some small peaks with (i) linear and (j) logarithmic scales.}	
\end{figure*}

Figure \ref{fig6} summarized the raw ARPES spectra used in the present paper. As for the results in Fig. \ref{fig2}, we extracted all the peak positions of the SS2 bands and those of the SS1 bands near \mb point from the energy distribution curves and Lorentzian fitting. The SS1 bands have sharp dispersions around \gb point and the peak positions were extracted by fitting the momentum distribution curves. One may notice in Figs. \ref{fig6}(a) and \ref{fig6}(c) that strange peaks showed up near \EF at $k=$ 0.2$\sim$0.4 \AA$^{-1}$ as highlighted by black arrows. We can confirm they do not correspond to the real band positions by tracking peak positions in the momentum profiles of Figs. \ref{fig6}(d) and \ref{fig6}(f). They seem to be generated by inelastic scattering processes near the Fermi level. A similar phenomenon seems to have occurred in the SS1 bands near \mb point in Figs. \ref{fig6}(b) and \ref{fig6}(c), where the peak profiles appear asymmetric and broadened. Considering the calculated surface band dispersions and the fact that no new ordered structure was observed by LEED at the present Cs coverages, the change in the peak profile does not mean the existence of new bands. To fit such asymmetric peaks, we just magnified the summit and applied Lorentzian fitting there. As for fitting of QWS peak positions in Figs. \ref{fig6}(g)-\ref{fig6}(j), empirical polynomial backgrounds were used. The procedures do not directly consider complicated background contributions, but we confirmed such details in the fitting process did not seriously affect the extracted peak positions and did not change the present conclusions.

\section{Appendix B: Reproducing the surface band deformation using a phase accumulation model}

\begin{figure}
\includegraphics{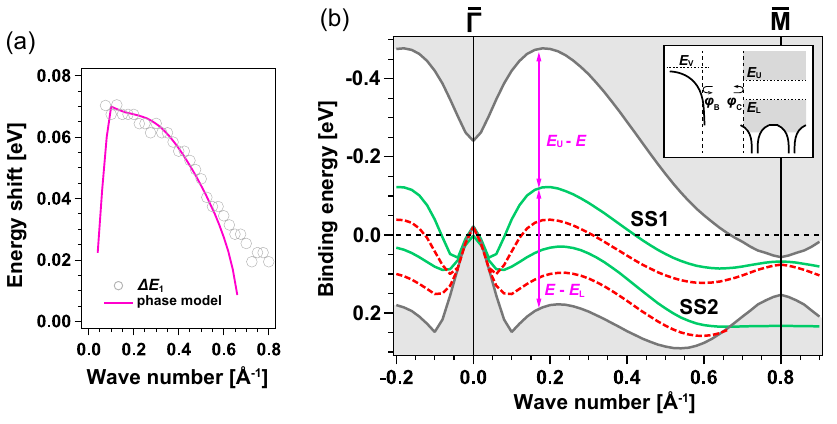}
\caption{\label{fig7}(a) Comparison of experimental SS2 band shifts [the same as in Fig. \ref{fig2}(e)] and the ones obtained from an analysis using a phase accumulation model. (b) Simulated surface band structures before (green solid line) and after (red dashed line) applying the band shifts obtained from the phase analysis. (inset) Schematic of a phase accumulation model for surface states.} 	
\end{figure}

Focusing on surface confinement potentials and phase shifts of confined wave functions, an effective model called a phase accumulation model has provided an analytical formulation to understand experimental results on surface states and QWSs \cite{hirahara2007, ito2016,chiang2000,smith1985,smith1994,neuhold1997,matsuda2002, davison1992,hufner2013}. For example, the condition for a surface state to exist is expressed as
\begin{equation}
	\phi_\mathrm{B}+\phi_\mathrm{C}=2\pi n
\end{equation}
where $n$ is a quantization number starting from zero and $\phi_\mathrm{B}$ and $\phi_\mathrm{C}$ show, respectively, phase shifts at a surface image potential and at a crystal surface potential [see the inset of Fig. \ref{fig7}(b)]. Although it is not explicitly included in the formula, a surface state has a complex-valued vertical momentum, whose imaginary part serves as a decaying term for the amplitude of the wave function \cite{davison1992, hufner2013}. As for the phase shifts, the following formulas are generally used \cite{smith1985,smith1994}:
\begin{align}
	\phi_\mathrm{B}&=\pi\sqrt{\frac{\mathrm{3.4 [eV]}}{E_\mathrm{V}-E}}-\pi \\
	\phi_\mathrm{C}&=2\arcsin\sqrt{{\frac{E-E_\mathrm{L}}{E_\mathrm{U}-E_\mathrm{L}}}}-\pi
\end{align}
$E_\mathrm{V}$ is the vacuum level and $E_\mathrm{U}$ and $E_\mathrm{L}$ are the upper and lower edges of the bulk band, respectively. We can obtain a relation between $E$ and $E_\mathrm{V}$ by substituting Eqs. (2) and (3) into Eq. (1), where a reduction of the work function owing to ionization of alkali-metal atoms results in a change of the energy position of the surface state. Since $\phi_\mathrm{C}$ depends on $E_\mathrm{U}$ and $E_\mathrm{L}$, the energy modulation can exhibit wave-number dependence via that of the bulk band-edge positions. The dependence can be explicitly formulated as
\begin{equation}
	\Delta E(\bm{k})=\frac{\Delta E_\mathrm{V}}{1+\frac{2}{\pi\sqrt{\mathrm{3.4 [eV]}}}\sqrt{\frac{(E_\mathrm{V}-E(\bm{k}))^3}{(E_\mathrm{U}(\bm{k})-E(\bm{k}))(E(\bm{k})-E_\mathrm{L}(\bm{k}))}}}
\end{equation}
The larger the energy separation between the bulk band edges and the surface band $(E_\mathrm{U}-E)(E-E_\mathrm{L})$ is, the larger the band shift becomes for a fixed $\Delta E_\mathrm{V}$ value. When the surface state is located far from the bulk band edges, the state is strongly confined by the crystal surface potential, whose boundary condition (phase shift $\phi_\mathrm{C}$) is not much affected by a perturbation to the system and, instead, the phase shift $\phi_\mathrm{B}$ at the surface potential becomes more sensitive to the perturbation, which leads to the larger energy shift.

Figure \ref{fig7}(a) compares the experimental band shifts for the SS2 band [the ones shown in Fig. \ref{fig2}(e)] with the ones calculated using Eq. (4). The values for $E_\mathrm{U}$ and $E_\mathrm{L}$ were extracted from Fig. \ref{fig1}(e) and the work function of Bi, that is $E_\mathrm{V}-E_\mathrm{F}$, was set to 4.36 eV \cite{lide2004}. The work function lowering $\Delta E_\mathrm{V}$ was treated as a fitting parameter and was adjusted to 0.85 eV to best fit the data. We also simulated the resulting band structures after applying the band shifts in Fig. \ref{fig7}(b). Although the overall tendency was nicely reproduced, finite deviations were confirmed at wave numbers larger than $\sim$0.6 \AA$^{-1}$ in the SS2 band shifts. One possibility is that we might need to consider a hybridization effect between the surface states and the projected bulk states. In addition, discontinuity of the boundary condition at the bulk band edge can modify the simple formula of Eq. (3) around the region \cite{matsuda2002}. Moreover, in first-principles calculations, the position of surface bands are sensitive to a structural relaxation around surfaces \cite{koroteev2008} and including the effects may improve the prediction. 

\bibliographystyle{apsrev4-1}
\bibliography{reference_CsBi.bib}

\end{document}